\documentclass{paper}       
\usepackage{graphicx}
\usepackage{amssymb,amsmath,multirow,rotating}
\usepackage[authoryear,round]{natbib} 

\newcommand*\mylabel[1]{\label{#1}}

\newcommand*\myref[1]{(\ref{#1})}
\begin{document}

\title{Phase Transition in the S\&P Stock Market
}

{\small
\author{Matthias Raddant\\
Institute for the World Economy, Kiellinie 66, 24105 Kiel, and \\
Institut f\"ur Volkswirtschaftslehre, Universit\"at Kiel, Germany\\
\itshape{matthias.raddant@ifw-kiel.de} 
 \and 
Friedrich Wagner \\
Institut f\"ur Theoretische Physik und Astrophysik\\ Leibnizstra{\ss}e 15, Universit\"at Kiel, Germany\\
							\itshape{wagner@theo-physik.uni-kiel.de}}
}

\maketitle

\vspace{0.1cm}
\begin{center}
{\itshape 
 The final publication is available at link.springer.com\\
http://link.springer.com/article/10.1007/s11403-015-0160-x\\
to be published in the Journal of Economic Interaction and Coordination.
}
\vspace{0.5cm}
\end{center}

\begin{abstract}
We analyze the returns of stocks contained in the Standard \& Poor's 500 index from 1987 until 2011. We use
 covariance matrices of the firms' returns determined in a time windows of several
years. We find that the eigenvector belonging to the leading eigenvalue (the market)
exhibits a phase transition. The market is in an ordered state from 1995 to 2005  and in a disordered state after 2005.
 We can relate this transition to an order parameter derived from 
the stocks' beta and the trading volume. This order parameter can also be interpreted within an agent-based model. 

\end{abstract}

\section{Introduction    \label{intr} }

In this paper we analyze the structure of the U.S. stock market. We show that the influence 
of stocks on the market is changing and that this influence is related to trading volume 
and the stocks' betas.

The analysis of the structure of stock markets is dominated by two research approaches.
The first one tries to explain the differences between the rates of return of stocks
and relates to the seminal work by \cite{lintner} and \cite{sharpe} and the CAPM model. The second approach takes the investor's point of view, and is hence mostly focused on the choice of a portfolio and the analysis of risk.
Both are related by the need to evaluate the comovement of stocks
with each other and some index or market proxy.

The original version of the CAPM is in fact a one-factor-model, which postulates that the
returns $r_i$ of the stocks should be governed by the market return $r_M$ and only differ by
the an idiosyncratic component $\beta_i$ for each stock $i$, such that
\begin{eqnarray}
r_i(t) = \alpha(t) + \beta_i r_{M}(t) + \epsilon_i(t) .
\end{eqnarray}
In this setting $\alpha$ can also be interpreted as the risk free rate of interest.
Hence, stocks differ by the amount of volatility with respect to the market, and economic
rationale necessitates that higher stock volatility is compensated by higher absolute returns.
Empirical tests of this model had rather mixed results and have let to the conclusion that 
beta values are not constant but time-varying, see \cite{tvcapm}. The \cite{famafrench} model
extends this approach to a three-factor model, incorporating firm size and book-to-market ratio.
 Several other extension of the original 
models have been suggested, mostly building on some kind of conditional CAPM, where the entire 
model follows a first-order auto-regressive process, see \cite{bodurtha}.
The reasons for the change of the betas are manifold. They could change due to microeconomic
factors, the business environment, macroeconomic factors, or due to changes of 
expectations, see, e.g., \cite{bos}. \cite{adcock}, \cite{harvey} and \cite{plerou} have discussed that the non-normality of stock returns, 
conditional skewness, and mainly the long memory in returns can lead to distorted estimations of the CAPM. There is also a strand
of literature that tries to capture the effects of heterogeneous beliefs of investors in a CAPM framework, e.g.,
\cite{brockhommes} and \cite{carl}.

In order to manage the risk of a portfolio, one can derive optimal portfolio weights from the 
spectral decomposition of the covariance matrix of stock returns. Many studies show that the non-normality 
of stock returns can lead to an under-estimation of risk. A common way to describe the properties 
of stock comovements is to look at the eigenvalue spectrum of the correlation matrix.
Random matrix theory suggests that a market that behaves like a one-factor model should
result in one dominant eigenvalue. Both, the non-normality in the data and any influence from other factors
will result in deviations from this simplified model, see \cite{cizeau} and \cite{alfa}. 

Approaches which utilize the spectral properties of correlations matrices have their limits
once the number of variables becomes large relative to the number of observations.
Networks approaches, which derive dependency networks from the correlation matrix can be useful,
as long as one does not need explicit portfolio weights for each single stock, see, e.g., \cite{dnetwork} and \cite{montegna}.
A related approach is to try to identify different states of the stock market, either by an analysis 
of the correlation matrix like in \cite{munnix}, or by the analysis of transaction volumes as in \cite{preis}.
Recent studies like \cite{raddant} show that the correlation structure in stock markets are rather volatile, and partly
mirror economic and political changes. \cite{kenn} for example shows that a structural break seems to 
happen in the U.S. market around 2001. This strand of literature is also related to approaches from
econometrics.  \cite{beile} for example argue that correlations increase in times of crisis, which has
profound implication for portfolio choice and hedging of risks. Other studies like \cite{ahlgren} analyze if correlations
in and between markets have increased due to more openness and tighter economic relations
between countries.

Since financial markets tend to react very fast on any change in the economy, but also inhibit a lot
of noise, we found that a look at longer time horizons is a worthwhile contribution to the field, since many
of the above mentioned studies look at time horizons of months or a few years. We found that the S\&P 500
contains around 170 stocks with a history of price quotes of 25 years (the number drops rapidly with much more than this 25 years).
We analyze the long-run development of the stocks influence upon the market. We derive both, 
a market index and the stocks' influence, from the spectral decomposition of the covariance matrix.
We show that for most of the period under consideration the market was in a ordered state, characterized by
a disproportionate influence of stocks from the IT sector. While some changes in the market seem to happen in 1995,
the collapse of this regime starts with the burst of the dot.com bubble. A disordered state is found after 2005. We
will show that from here the market develops into a new (although weaker) ordered state, where the driving sector is
the financial industry. 

The paper is organized in the following way. In section \ref{data} we describe
the subset of stocks in the S\&P market used in this analysis. Section \ref{corr}
contains the definition of market indices derived from the covariance matrix.
In section \ref{size} we explain why we prefer the latter over the usually applied correlation
matrix, and that the average return $r_{av}$ and market return $r_M(t)$
may be almost equal when calculated from a large number of stocks. Section
\ref{phas} contains our results for the phase transition and section \ref{conc}
some conclusions.

\section{Materials and Methods \label{data} }

The most important criterion for data selection consists in the length of
the time series $T$ of stock prices. Only 289 firms in the S\&P 500 are listed from
 from January 1987 until December 2011, but not all of them will qualify for our analysis.
 In the present work we study the covariance 
matrix of firm returns. To a large part this matrix is a random matrix, where
the errors of the quantities of interest are in the order of
$\sqrt{N/T}$, see also \cite{marc}. Therefore one can in fact afford a reduction of $N$
by the following criteria:
We start out with the 500 stocks which are listed as part of the S\&P500 index 
at the end of 2011. We drop all those which were not trading since January 1987.
We then filter for illiquid stocks; we define stocks as illiquid if their price does
not change for more than 7\% of the trading days. We validate this selection by 
checking the daily trading volumes. Further we delete single stocks which price does not
move for at least 10 days in a row (e.g. due to suspended trading).

Our final set of data comprises the stock prices of $N=171$ firms at 
$T=6312$ trading days in the time window 1987-2011. As a sign of the 
different sizes of the firms we will later also consider the yearly trading volume
of the firms in this period.  A disadvantage of our selection
consists in the loss of the meaning of the index, since this refers to
 a changing set of 500 firms and may not be representative for our subset.
This selection also leads to some form of selection bias, since failed enterprises
are excluded from the analysis.

A frequently used tool to analyze financial markets consists in the study of the
correlation matrix between the stock returns of a market. 
This matrix can be used in two ways. Its observation needs a certain time 
window $t_W$. For small window sizes (10-20 days in case of daily returns),
the matrix is dominated by noise and a principal component analysis does not
make any sense. In the first class of studies like \cite{borl} and \cite{kenn}, the noise is
reduced by averaging the correlation matrix over the stocks. This means to replace
the volatility of the average return $r_{av}$ by the sum of firm returns.

On the other side, when choosing $t_W$ in the order of a few years, the decomposition
into eigenvectors may be meaningful. The correlation matrix possesses one large
eigenvalue in the order of the number $N$ of stocks. The corresponding
eigenvector can be used as a description of the market, see \cite{lalo,galu}.
The remaining eigenvalues are qualitatively similar to those of a random 
matrix with a \cite{marc} spectrum. Nevertheless, with the
assumption of a specific model, information can be extracted from this part of the
spectrum, see \cite{alfa}. In general, only eigenvalues separated by more than
$\sqrt{N/t_W}$ from other values have a model independent meaning, see \cite{burd}.
Therefore we concentrate on the long term time behavior
of the eigenvector of the market eigenvalue.\footnote{Extending the window too much
could in fact at some point lead to problems, even if one assumes that the betas are
slowly varying, see for example \cite{livan}.}

A daily market return $r_M(t)$ can be obtained by the scalar product of the 
stock returns with the eigenvector, determined in an appropriate window. In this
case the eigenvector denoted by $\beta_i$ (normalized to $\sum_i \beta_i^2=N $) describes the $\beta$ coefficients
relative to the market, as needed for a CAPM portfolio in the style of \cite{capm}. This interpretation
is supported by the empirical results of \cite{lalo}, the $\beta_i$ are positive and
and distributed around one for $t_w>3$y. 

In an alternative description of the market, the average return $r_{av}$ 
may be used, see \cite{borl} and \cite{kenn}. First, we investigate how much $r_{av},r_M$ and the
index return differ from each other. Secondly we will investigate the time
dependence of $\beta_i$ to find evidence for a phase transition.

Transitions in physics are characterized by an order parameter $m$, which vanishes in the disordered phase and
is non-zero in the ordered phase. $m$ could be related to macroscopic or microscopic properties
of the system. In general, $m$ is discontinuous at the critical point (first order). In special
cases, the transition is of continuous order with continuous $m$. Corresponding models of
statistical physics near the critical region 
have been applied to financial markets in \cite{bouc,stau,born}. They offer an explanation of the 
stylized facts of the returns, see \cite{tlux}. However, due to the universality, the relation 
of the model parameters to economical quantities remains obscure. The models 
require fine tuning of the parameters to maintain the system close to the critical 
region. Since these systems always stay in 
a disordered phase, neither a micro- nor macroeconomic order parameter can be
 observed directly.
In this study we look for a first order transition of stocks that are characterized
by high volatility ($\beta_i >1$) and a high trading volume, which we capture by calculating
 a macroeconomic order parameter $m$.

\section{Correlation Matrix and Pseudo Indices \label{corr} }
\subsection{Properties of the correlation matrix}

The daily stock stock prices $S_i(t)$ for stock $i=1,\ldots,N$ at day $t=1,\ldots,T$ may be
converted into returns $r_i(t)$ by
\begin{equation} \mylabel{21}
r_i(t)=r_N\ln(S_i(t+1)/S_i(t))
\end{equation}
We use a normalization factor $r_N$ given by $\sum_{i,t}r_i^2(t)=NT$.
To see the time dependence of the correlation between stocks we construct the
covariance matrix $C$ at time $\tau$
 by selecting a time window\footnote{In this paper $t$ is used to specific days, 
while $\tau$ is used to index variables like $C$ that are calculated for a time window.}  of size $t_w$ 
centered at $\tau$
 
\begin{equation} \mylabel{22}
C_{ij}(\tau)=\frac{1}{t_w}\sum_{t=\tau-t_w/2}^{t=\tau+t_w/2}\; r_i(t)r_j(t)
\end{equation} \label{eq:corr}
Many  previous investigations (as discussed in the introduction) use the Pearson correlation 
\citep{Pearson}, which describes the covariance between standardized returns.
Subtracting from each time series the mean of $r_i$ would be a small effect, but setting the variance to 1 may mask
a possible $\tau$ dependence of the eigenvalues of $C(\tau)$. For large window sizes
$C$ has one large eigenvalue of order $N/3$ and the corresponding eigenvector has only
positive components, see, e.g., \cite{plerou}. Empirically we found that these properties are lost for
window sizes less than 3 years. Therefore we will in the following use a window sizes $t_w = 4$ years. 
To compensate for the loss of time resolution
we use overlapping bins by choosing time steps in $\tau$  less than $t_w$.

The matrix $C$ can be written in a spectral decomposition as
\begin{equation} \mylabel{24}
 C_{ij}(\tau)=\sum_{\nu=0}^{N-1}e_i^{\nu}(\tau)\;
                                e_j^{\nu}(\tau)\lambda_\nu(\tau)
\end{equation}
where $\lambda_\nu$ denotes the eigenvalues and $e_i^{\nu}$ the eigenvectors
of $C$. The large eigenvalue corresponding to $\nu=0$ and its eigenvector 
$e_i^{0}$ can be interpreted as a description of the market, see, e.g., \cite{plerou}.   
It can be used to define a market return $r_M(t)$ for $t$ in a window
\begin{equation} \mylabel{25}
r_M(t)=\frac{1}{\sqrt{N}}\sum_{i}\; e_i^{0}(\tau)r_i(t)
\end{equation}
This means that the market return (which is also a market index) consists of the weighted contributions of
the returns of the single stocks. The weights are proportional to the entries 
in the eigenvector that corresponds to the leading eigenvalue.

For the CAPM model, one needs so-called $\beta_i$-coefficients, which
describe how close a stock follows the market described by a reference
return $\bar{r}$.
\begin{equation} \mylabel{26}
\beta_i=\frac{E[r_i\; \bar{r}]}{E[\bar{r}^2]}
\end{equation}
In a dynamic analysis the vector of beta values becomes time dependent. For our analysis we calculate 
betas in time windows $\tau$. As a reference return we use 
the market return $r_M$ instead of the real S\&P-index,
because the latter may not be representative for our data selection.
Replacing the expectation values in equation \ref{26} by taking time averages we obtain
with  $\bar{r}=r_M$ from equation \ref{24} the following $\beta_i$ coefficients
\begin{equation} \mylabel{27}
\beta_i(\tau)=\sqrt{N}e_i^{0}(\tau)
\end{equation}
After resolving the sign ambiguity in $e_i^{\nu}$ we recover the fact
that for $t_w \ge 3 $ years 
all $\beta_i$ are positive. Due to the normalization $\sum_i(e_i^{0})^2=1$
the $\beta_i(\tau)$ are distributed around a mean close to 1.
Firms with large $\beta_i$ follow the market more than others and also
they influence the market more than firms with small $\beta_i$.
For this reason we will refer to stocks with $\beta_i$ larger than a given threshold $\beta_c$ as the 
\emph{market leaders}.

\subsection{The market}

\begin{figure}[ht]
\centering
   \includegraphics[width=0.9 \linewidth, bb=29 367 568 634, clip=true]{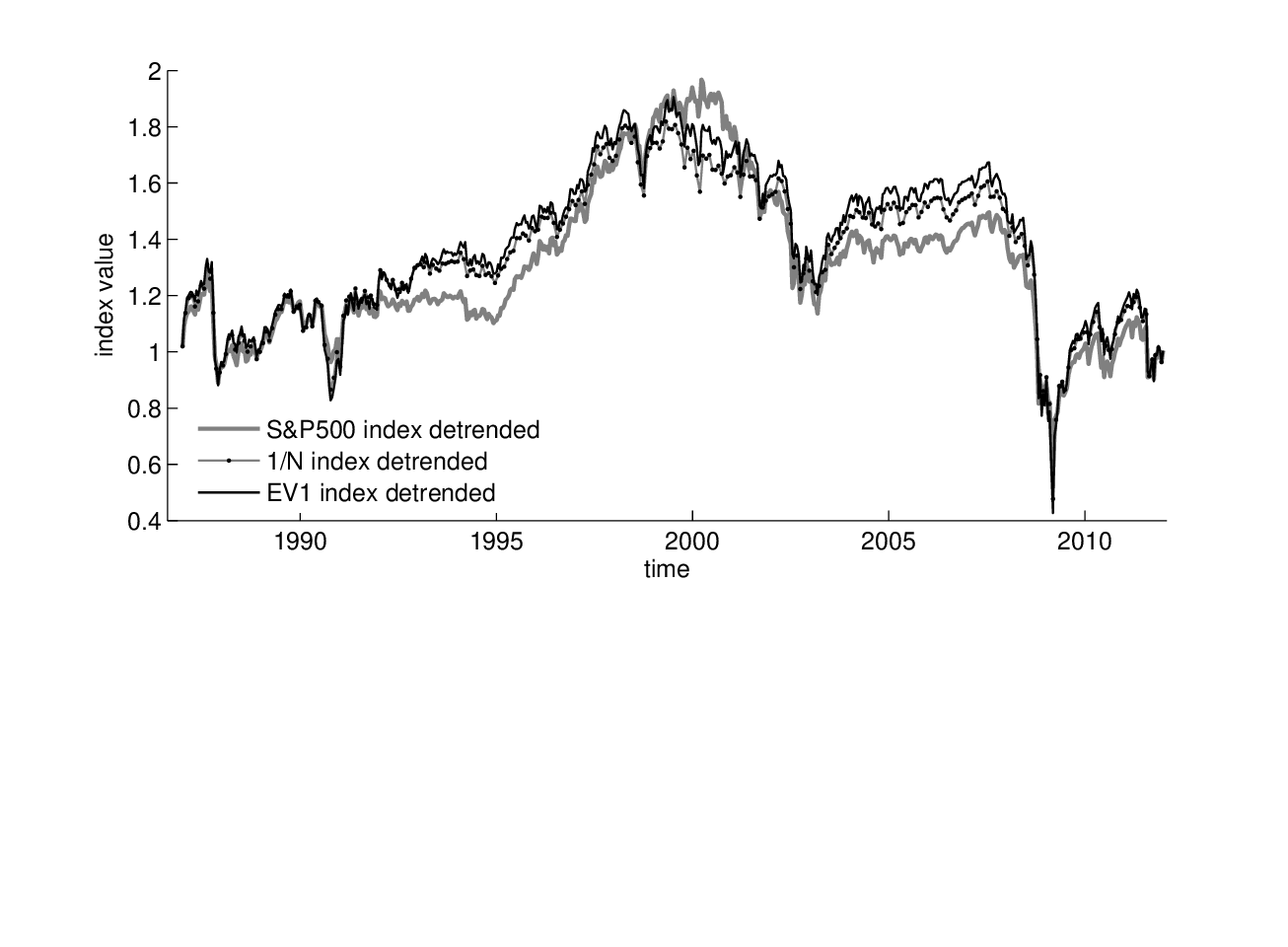}
   \caption{\label{fig1}{\em S\&P500 index (grey), and pseudo indices. $L_{av}$ (solid)
	and $L_M$ (dotted) are very similar. They show changing deviations from the real S\&P index.}}
\end{figure}

A simple way of defining a market would be the average return
\begin{equation} \mylabel{28}
r_{av}(t)=\frac{1}{N}\sum_{i}\; r_i(t).
\end{equation}
$ r_{av}^2$ averaged over time corresponds to a correlation matrix, averaged
over the stocks, which has also its defenders in the literature, see \cite{borl,kenn}.
From the returns defined by equation \myref{28} we can calculate
logarithms of a pseudo index $L_{av}$  by the recursion 
\begin{equation} \mylabel{29}
L_{av}(t+1)=L_{av}(t)+\frac{r_{av}(t)}{r_N}.
\end{equation}
The market return $r_M(t)$ from equation \myref{25} refers to a specific window.
To define a market pseudo index $L_{M}$ 
we use $\bar{r}_M(t)$ which is defined for all $t$ by equation \myref{25}
with $e_i^{0}(\tau)$ of the window with center $\tau$ next to $t$.
\begin{equation} \mylabel{29aa}
L_{M}(t+1)=L_{M}(t)+\frac{\bar{r}_{M}(t)}{r_N}.
\end{equation}
The recursions \myref{29} and \myref{29aa} can be integrated.  
The integration constants are fixed by the normalization
\begin{equation} \mylabel{29a}
\sum_t\; L_{av}(t)=\sum_t\;L_{M}(t)=0
\end{equation}
We compare these pseudo indices in figure \ref{fig1} with the 
index $S_0(t)$, written in the form $L_{0}(t)=\ln S_0(t) -l_0$. 
$l_0$ ensures equation \myref{29a} also for $L_0$.
As expected, the market index $L_{M}$  calculated with a time window of
$t_w=4$ years agrees with $L_{0}$ only qualitatively.
The average pseudo index $L_{av}(t)$ is very similar to the market pseudo index
$L_{M}$. This is somewhat surprising since we will see 
that $e_i^{(0)}(\tau)$ can change with time.
Especially in the years 2001-2008 the index exhibits considerably
larger average returns than the pseudo indices. This effect can be interpreted as 
phase transition into a stiff market. It was detected in \cite{kenn}, 
where correlations with the index have been subtracted from the
stock correlations. An increase of the former correlation leads to smaller
subtracted correlations after 2001. 
 For the squared difference $\Delta^2$ of $r_M$ and $r_{av}$ given by
\begin{equation} \mylabel{29b}
\Delta^2(\tau)=\left[\sum_{t=\tau-t_w/2}^{\tau+t_w/2}(r_M(t)-r_{av}(t))^2\right]\;
                    \left[\sum_{t=\tau-t_w/2}^{\tau+t_w/2}r_M^2(t)\right]^{-1}
\end{equation}
we derive in appendix A the inequality
\begin{equation} \mylabel{29c}
\Delta^2(\tau) \le (1-\bar{\beta})(\frac{2}{\lambda_0}\cdot trace(C)\;-1-\bar{\beta})
\end{equation}
with $\bar{\beta}$ the mean of $\beta_i$. The average correlation $C_{av}$
\begin{equation} \mylabel{29d}
C_{av}=\frac{1}{N(N-1)}\sum_{i \ne j}C_{ij}
\end{equation}
can be expressed by $\Delta^2$ and the properties of the market 
component (see appendix A) up to terms of order $1/N$
\begin{equation} \mylabel{29e}
C_{av}=<r_M^2>(\Delta^2+2\bar{\beta}-1)-\frac{1}{N}
\end{equation}

In the next section we discuss that
for large markets ($N \to \infty$) both, empirically and in context of a
model, $\bar{\beta}$ approaches one and therefore $\Delta^2$ vanishes.
In this limit $C_{av}$ corresponds to the volatility $<r_M^2>$ of the market.

\section{Dependence on the Market Size \label{size}}

The qualitative behavior of the correlation matrix suggests a decomposition 
of the returns $r_i$ according to a stochastic volatility model, see \cite{stvm}. 
$r_i(t)$ is the sum of the two
products, of noise $\eta$ and the market and of noise and the remaining contribution.
The coefficients $\beta$, the coupling $\gamma_M$ to the market, and the 
idiosyncratic couplings $\gamma_i$ are assumed to be constant in each window.
\begin{equation} \mylabel{31}
r_i(t)=\beta^0_i\;\gamma_M \cdot \eta_m  + \gamma_i \cdot \eta_i
\end{equation}
The independent noise factors $\eta$ have mean 0 and variance 1. 
The coefficients $\beta_i^0$ are normalized to $\sum_i (\beta_i^0)^2=N$.
For $t_w \to \infty$ $C$ can
be determined from the expectation values
\begin{equation} \mylabel{32}
C_{ij}=\beta_i^0\beta_j^0\gamma_M^2 + \delta_{ij}\; \gamma_i^2
\end{equation}
For large $N$ one can solve the eigenvalue problem for $C$ by a 
$1/N$ expansion (see appendix B). $C$ has one large eigenvalue
\begin{equation} \mylabel{33}
\lambda_0=N\gamma_M^2 +\langle\gamma^2\rangle_\beta +
         (\langle\gamma^4\rangle_\beta-\langle\gamma^2\rangle_\beta^2)/(N\gamma_M^2)
\end{equation}
with an eigenvector corresponding to
\begin{equation} \mylabel{34}
\beta_i=\left [1+(\gamma_i^2-\langle\gamma^2\rangle_\beta)/(N\gamma_M^2) \right ]\beta_i^0
\end{equation}
$\langle a \rangle_\beta$ denotes an average over $a_i(\beta_i^0)^2$. 

\begin{figure}[ht]
\begin{center}
   \includegraphics[width= 0.9\linewidth]{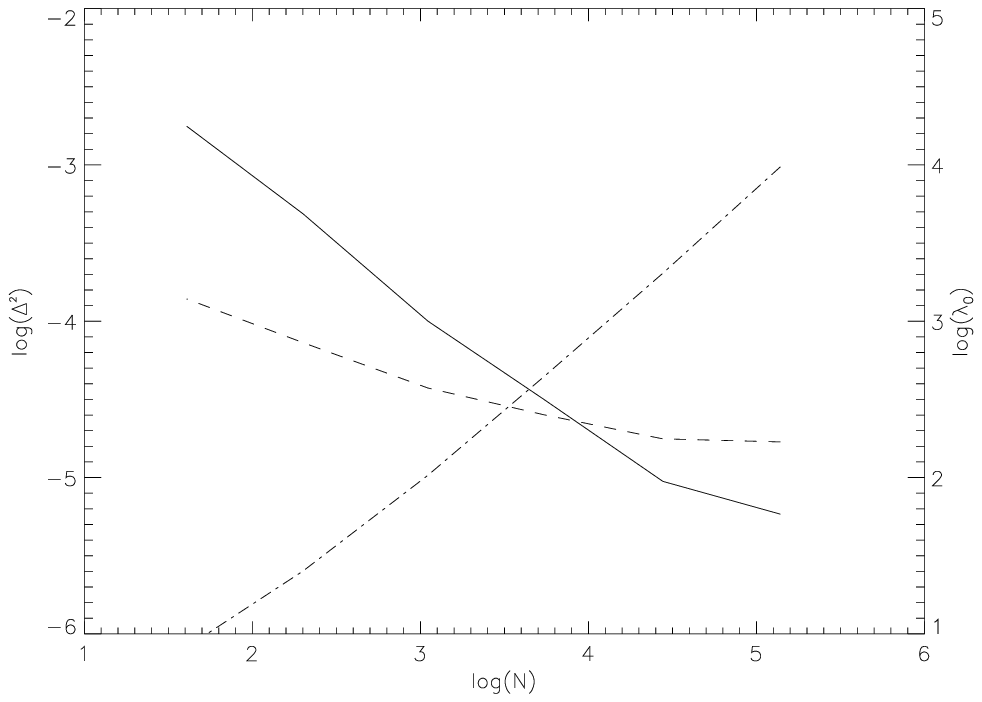}
\end{center}	
    \caption{\label{fig2}{ \em The solid line shows the squared difference
       between market and average return as $\ln(\Delta^2(N))$ as
       function of $\ln N$ (left hand scale). The dashed dotted line gives
       $\ln(\lambda_0)$ as function of $\ln N$ (right hand scale) and the
       dashed line the variance of the distribution of $\beta_i$ on a
       logarithmic scale.} }
			
\end{figure}

Neglected terms in equations \myref{33} and \myref{34}
are of order $1/N^2$. Empirically the $\beta_i$ are distributed around a mean
$\bar{\beta}$ close to 1 and a variance $\sigma_{\beta}$ decreasing with $N$.
Therefor, we can assume the model parameter $\beta_i^0$ to be equal to one.
From the inequality \myref{29c} we see that the difference between $r_M$ 
and $r_{av}$ expressed by $\Delta^2$
(see equation \myref{29b}) has to vanish in this limit.

In the following we investigate the behavior of the leading eigenvalue 
$\lambda_0 \propto N$, $\sigma_{\beta}\propto N^{-\gamma_1}$ and 
$\Delta^2 \propto N^{-\gamma_2}$ as function of $N$.

 A finite window size
$t_w$ leads to deviations of the observed $C$ from the ideal $C$ of equation \myref{32} in
the order of $\sqrt{N/t_w}$. To minimize this systematic error, we choose
the maximum window size $t_w=T$. To have a varying $N$ we adopt the 
following procedure: Sub-markets are defined by dividing the $N_0$ stocks
into $k$ groups with size $N(k)=N_0/k$ such that each group contains the 
same fraction of large and small firms as the full set. To improve the 
statistics we average $\lambda_0$, $\Delta^2$ and $\sigma^2_{\gamma}$
over the groups. The result is presented in figure \ref{fig2}, which shows
these quantities as a function of $N(k)$ on a log-log scale. $\lambda_0$
(dashed dotted line) increases with $N^\alpha$ with a power of $\alpha$ close to 1.
$\Delta^2(N)$ (solid line) and $\sigma_{\gamma}(N)$ (dashed line)
exhibit a lesser decrease than the
expected $1/N$ behavior. Since they also flatten out at large $N$, this
indicates an influence of the finite observation time window. 
The observed values of $\Delta^2$ are much smaller than the
upper limit from the Schwartz inequality in equation \myref{29c}.
This indicates that the eigenvectors $e_i^{\nu}$ for $\nu >0$
are almost orthogonal to a constant vector $e_i=1/\sqrt{N}$ already at
finite $N$, which implies equality of $r_M$ and $r_{av}$.

To summarize, the data support a stochastic volatility model of a sum
of market and preferences $\gamma_i$ for individual stocks where for large
$N$ the stock returns couple to the market component in the same way.
Therefore $r_{av}$ can be taken as a description of the market. Since it
determines the average correlation $C_{av}$, the frequent
use of $C_{av}$ in the literature as a proxy for an index is empirically successful. 
Especially $r_{av}^2$ consists
in a good estimator for the leading eigenvalue $\sqrt{\lambda_0)/N}$, since
due to the law of large numbers the statistical error decreases with 
both, $\sqrt{N}$ and $\sqrt{T}$. From the perturbation expansion given in
appendix B we see that the leading eigenvalue and its vector can be more 
accurately determined than the remaining ones.

\section{\label{phas}Time Dependence of $\beta_i$ and Phase Transition}

In the previous section the coefficients $\gamma$
and $\beta$ have been discussed for a long time scale. In general however, 
 they will be time dependent. To minimize the influence of the noise
 produced by a finite observation window $t_w$, we chose a rather
large window of $t_w=4$ years. This implies that only 
long term changes in $\beta_i$ can be detected. In the following, we diagonalize $C$ for the S\&P data in overlapping
steps of 1 year.

\begin{figure}[ht]
\begin{center}
\begin{minipage}[t]{0.48\textwidth}
\centering
   \includegraphics[width= \textwidth, clip=true]{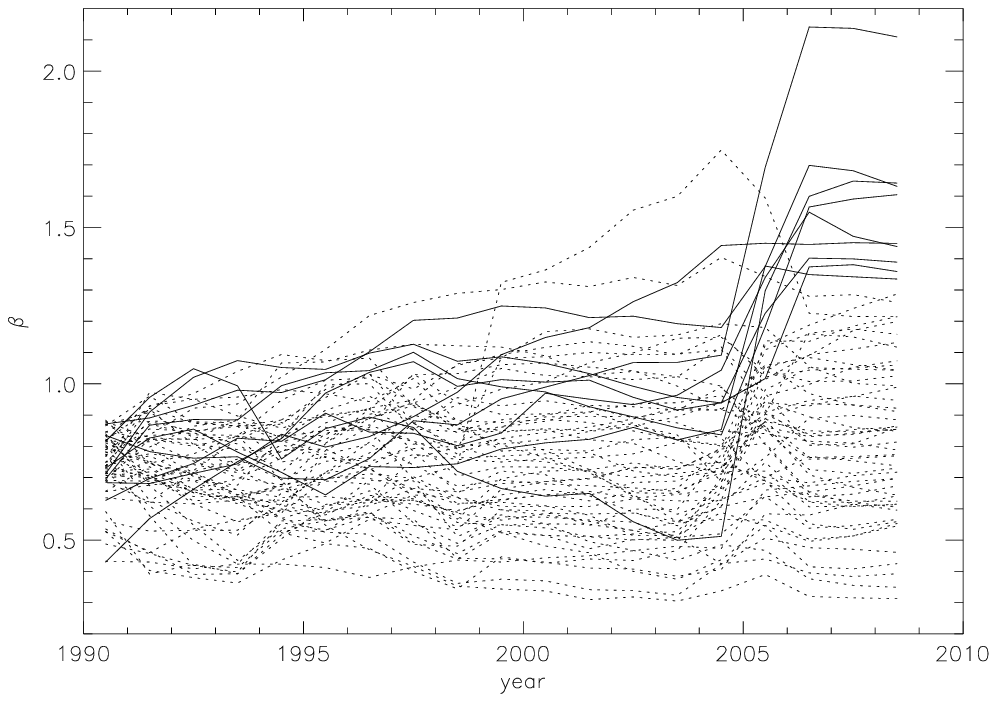}
   \caption{\label{fig4}{\em Time dependence of $\beta_i(\tau)$ for the 60
            firms with smallest trading volume. Solid lines indicate market
            leaders ($\beta_i>1.3$) in 2008.}}
 
\end{minipage}
\hspace{0.02 \textwidth}
\begin{minipage}[t]{0.48\textwidth}
\centering
   \includegraphics[width=\textwidth, clip=true]{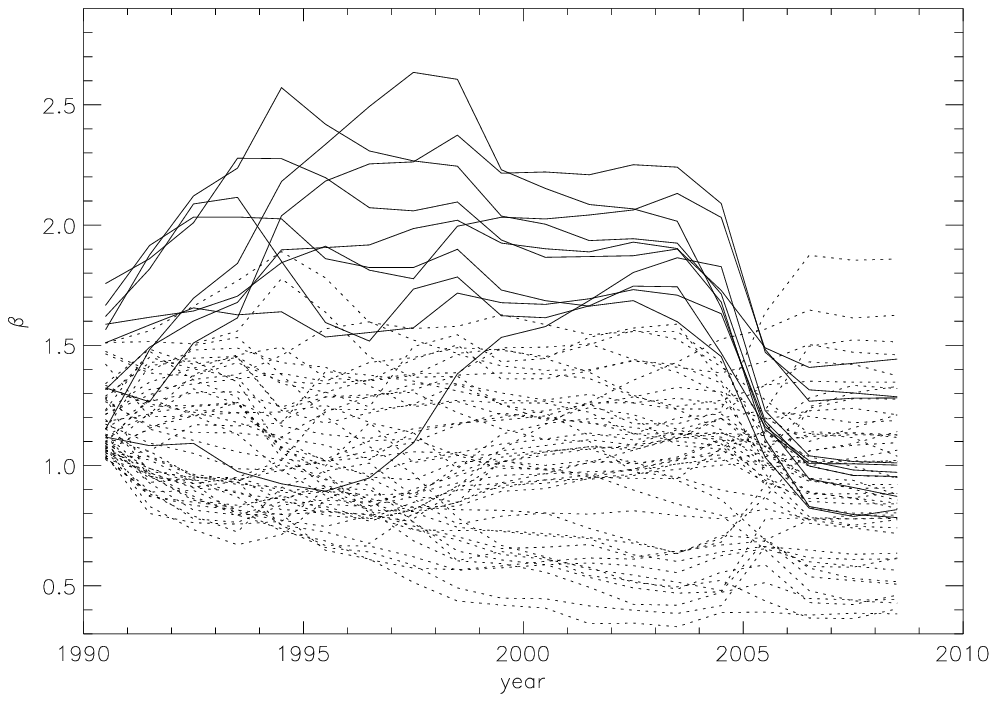}
   \caption{\label{fig3}{\em Time dependence of $\beta_i(\tau)$ for the 60
            firms with largest trading volume. Solid lines indicate market
            leaders ($\beta_i>1.3$) in 2002.}}
\end{minipage}
\end{center}
\end{figure}

The resulting time dependence of $\beta_i(\tau)$  is shown
in figure \ref{fig4} for the 60 firms with the smallest average traded volume.
Before 2005, their beta values are relatively constant with values $\le 1$, as expected
for small firms with less impact on the market. 
Around 2005 some firms with previously small $\beta$
experience a drastic increase and become market leaders (firms with
$\beta_i(2008) > 1.3$ are indicated by solid lines).
A different picture appears if we look at $\beta_i(\tau)$ for firms with large
average traded volume shown in figure \ref{fig3} for the 60 largest firms. Of
course this set contains market leaders. Those in 2002 are denoted by solid 
lines. However, in 2005 they disappear in favor of new firms as in the case 
of small firms.  Therefore in 2005 a reorganization of the market has happened.
This depends on the the sector classification of the firms.
The 18 market leaders in the set of all firms with $\beta_i>1.39$ in the
year 2002 are listed in table \ref{tab1} and in table \ref{tab2} those for the year  
2008. The 2002 list contains dominantly large firms from the
IT sector. After the transition in 2008 the list contains
firms of all sizes spread over many sectors.

\begin{table}
\begin{center}
\footnotesize
  \begin{tabular}{|c c| c c|} \hline
    Sector name &  \# in sample &  Sector name &  \# in sample\\ \hline
Consumer Discretionary & 24 & Consumer Staples       & 22 \\
Energy                 & 10 &Financials             & 28 \\
Health Care            & 18 & Industrials            & 33 \\
IT                     & 20  & Materials              & 14 \\
Telecommunication Services & 3 & Utilities              & 0 \\
\hline
 \end{tabular}
\end{center}
 \caption{\label{tab3}{{\em  List of} S\&P {\em sectors and frequency in the 
           data set, GICS classification}}}
\end{table}

This behavior of $\beta_i$ might indicate a phase transition.  A
macroeconomic order parameter that explains this transition 
should take into account the (large) $\beta_i$ 
values, the traded volume (see also \citep{dark}) and the sectors $s$ of the firms.  For $s$ we use the
GICS classification scheme into $S=10$ sectors given in table \ref{tab3}.
From the point of view of an investor, one can say that the
following function $R$ describes the risk (volatility) in each sector $s$
due to the $\beta$ coefficients
\begin{equation} \mylabel{40a}
R(\tau,s)=\sum_{i \epsilon s}\theta(\beta_i-1.0)\beta_i(\tau)\; v_i(\tau)
\end{equation}
In an ordered state one specific $R(\tau,s_0)$ is large and the remaining $R$'s are small.
A macroeconomic order parameter $m$ of a Potts model type can be obtained by normalizing $R$
\begin{equation} \mylabel{40}
m(\tau,s_0)=\frac{S}{S-1}\left [\frac{R(\tau,s_0)} {\sum_{s'}R(\tau,s')} 
                              \; - \frac{1}{S} \right ]
\end{equation}
If the appearance of one large $R$ is due to a transition to an ordered state, $m(\tau,s_0)$
will be close to 1 and the 'wrong' order parameters $m$ for $s\ne s_0$ will scatter
around small negative values due to the analogue of thermal fluctuations. 
In the disordered phase all $m$ will fluctuate around 0.
\begin{figure}[ht]
\begin{center}
  \includegraphics[width=0.95\textwidth]{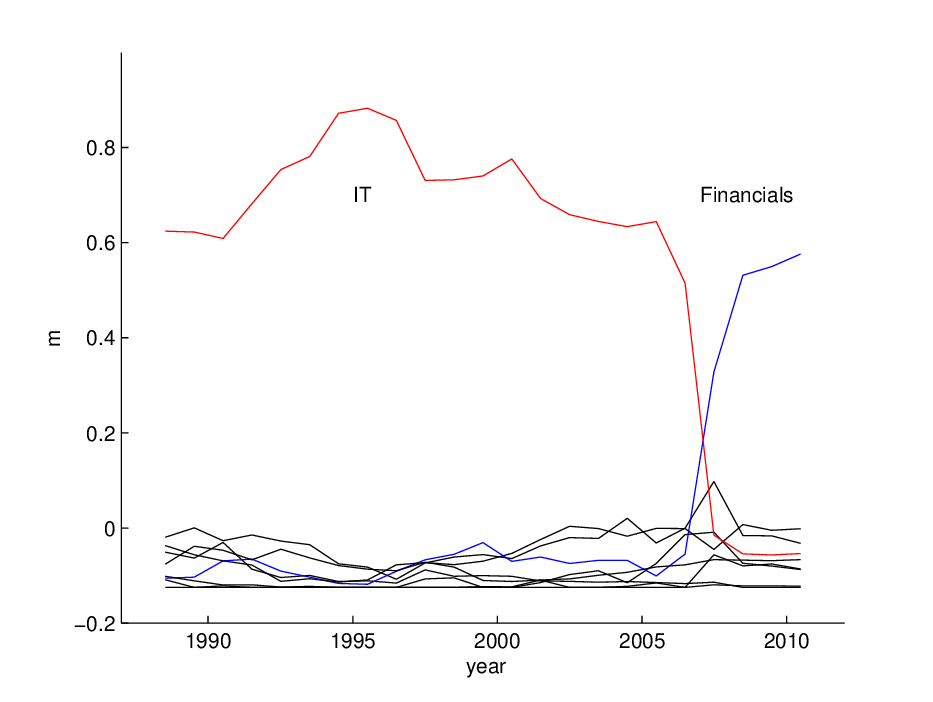}
   \caption{\label{fig5}\em The order parameters $m(\tau,s_0)$ for all sectors $s_0$
            as function of the center $\tau$ of the windows with size $t_W=4$. In the
            years 1992-2005 $m$ is large for the IT sector and small for
            the remaining sectors. Around 2010 the financial sector may give rise
            to an ordering.}
\end{center}
\end{figure}
In figure \ref{fig5} we show the order parameters $m(\tau,s)$ in time steps of
one year. Clearly $m$ is large for the IT sector, whereas all others are small negative.
After 2006 $m(\tau,IT)$  decreases.
Near the end of our time series an ordering towards the financial
sector may be possible.
\footnote{The dominance of the IT sector
and the change around 2005 can also be found for smaller window sizes of down to 2 years,
revealing one sharp peak around 1995/96. For much smaller time windows the 
influence of single events like the 1987 stock market crash or the 9/11 
attacks become rather large and distort long-run trends.}
Following the advice of an unknown referee we discuss in appendix \ref{randomize} that
the effect seen in figure \ref{fig5} is not due to the change of only a few firms.

\begin{table}
\footnotesize
 \begin{center}
  \begin{tabular*}{\textwidth}{|c @{\extracolsep{\fill}} ccc|cccc|}   \hline
     Firm &  Sector & $\beta$ & Vol. & Firm &  Sector & $\beta$ & Vol. \\ \hline
   TEXAS I. & IT         & 2.21 & 2966  &     HALLIBURTON & Energy     & 2.09 & 2641  \\ 
     ALTRIA & Cons. S.   & 2.04 & 8927  &      BANK OF A. & Finance & 1.94 & 3254  \\ 
  HEWLETT-P.& IT         & 1.89 & 2883  &       MICROSOFT & IT         & 1.87 & 19418  \\ 
 APPLIED M. & IT         & 1.69 & 6735 &         AM. EXP.& Finance & 1.69 & 1478  \\ 
     ORACLE & IT         & 1.66 & 11797 &           INTEL & IT         & 1.66 & 14429 \\ 
     PFIZER & Health     & 1.54 & 4272  &        ADOBE    & IT         & 1.53 & 1886  \\ 
      LOWE'S & Cons. D.   & 1.51 & 2511   &   JOHNSON \& J.& Health    & 1.50 & 2079  \\ 
    HERSHEY & Cons. S.   & 1.44 & 422 &            MERCK  & Health     & 1.40 & 1884  \\ 
      APPLE & IT         & 1.39 & 18540 &       INTERN.BUS.& Industry & 1.39 & 2347  \\ \hline
  \end{tabular*}
  \caption{\label{tab1}{\em  List of market leaders with $\beta(2002) \ge 1.39$.
           $Vol.$ gives the annual stock trading volume in millions of shares.}}
        \vspace{0.5cm}

  \begin{tabular*}{\textwidth}{|c @{\extracolsep{\fill}} ccc|cccc|}   \hline
   Firm &  Sector & $\beta$ & Vol. & Firm &  Sector & $\beta$ & Vol. \\ \hline
     HUMANA  & Health  & 2.11 &  828  &   INTERN.BUS.& Industry &  1.86 & 2379  \\ 
WEYERH.  & Finance & 1.86 &  1623 &         CHUBB & Finance  &  1.80 &  914  \\ 
VARIAN  & Health  & 1.64 &   367 &       TEXTRON & Industry & 1.63  &  873  \\ 
     CHEVRON & Energy  & 1.63 & 3923  & DONNELLEY &Industry & 1.60 &  412  \\ 
  A. DATA& Industry &1.55 &  823  &         TARGET &Cons. D.&  1.51 & 3191  \\ 
     LOWE'S   & Cons. D. &1.46 & 3893  &      C R BARD & Health  &  1.45 & 211  \\ 
       AVON  & Cons. S. &1.45 &  1098  &     MICROSOFT & IT       & 1.44 & 21314  \\ 
    G. PARTS & Cons. D. &1.44 &   309 &     PROG. OHIO&  Finance & 1.41 & 1461  \\ 
       CIGNA & Health   &1.39 &  918  &     WASH. PST.&  Cons. D.&1.39 &   8  \\ \hline
  \end{tabular*}
  \caption{\label{tab2} {\em  List of market leaders with $\beta(2008) \ge 1.39$.
           $Vol.$ gives the annual stock turnover in millions of shares.}}
 \end{center} 
\end{table}

There are different possibilities to relate this phase transition to the behavior of agents that trade in this market.
A microeconomic order parameter could be obtained by incorporating a behavioral agent-based model for the market
of the \cite{kirm} type. The sector specific returns could be defined as
\begin{equation} \mylabel{41}
r_s(t)=\theta_s(\tau) \cdot \eta_s(t) ,
\end{equation}
like in \cite{alfw} or \cite{frw1}. $\theta_s$ is proportional to the ratio of noise traders and fundamentalist 
agents that trade stocks in sector $s$. The i.i.d. Gaussian noise $\eta_s$ describes the noise traders.
$\theta_s$ is (on a longer time scale) time dependent, because the opinion of the agents changes.
 With suitable choice of the
parameters in the asymmetric Kirman model (see \cite{alfw} using the bimodal version)
this ratio can occasionally be large, and the system stays in this state
for longer times. Such a
situation cannot be distinguished empirically from a real phase transition.
A microeconomic order parameter related to the herding effect would then be given by
\begin{equation} \mylabel{42}
m(\tau,s)=\frac{\theta_s(\tau)}{1+\theta_s(\tau)}
\end{equation}

An alternative model is the application of a Pott's model with $S$ states, see \cite{Pott}.
The $\beta$ dependent interaction between agents is attractive if neighboring 
agents trade in the same sector. For strong enough dependent interactions the system
will order, with one sector dominating the others.

\section{Conclusions\label{conc} }

Our analysis of the market indices revealed that for a sufficiently large samples of stocks and longer time horizons
weighted indices differ only very slightly from any form of market average. This is a result of strong overall stock correlations
and relatively stable long-run correlation structures that we have shown by the analysis of the properties of the correlations matrix.

In our analysis of how the market is influenced by different stocks in the long-run, we have seen that the IT sector has played a dominant role
for a long time.
The time dependence of the CAPM coefficients $\beta$ exhibit a transition
in 2005. This is connected with the disappearance of a macroeconomic
order parameter for the IT sector. Despite our poor time resolution due
to the window size, this phase transition appears to be  sharp. The transition
 lies between the crash of the dot.com bubble and the
Lehmann desaster in 2008. In this time period we see no other pronounced effect in the
index or in the stock prices. 

A possible reason for a sharp transition
may be the following:
From 1990 to 2002 the stock prices experienced a steady increase. This
led investors to buy in the most increasing sector, the IT sector.
They minimized the risk by choosing only large firms. Disappointed
by the crash of the dot.com bubble in 2001, they changed their investment
strategy completely. Investments and trading volume became much more scattered over all segments. Figure
\ref{fig5} shows that later a (weaker) form of ordering took place by focusing on the financial sector.

\section*{Appendix}
\appendix

\begin{appendix}
\section{\label{app1} Correlation Matrix}
Denoting the time average as in equation \myref{22} by $[\;]_{\tau,t_W}$ we get from
equations \myref{24} and \myref{25} the average of $r_M^2$
\begin{equation} \mylabel{a1}
[r_M^2]_{\tau,t_W}=\frac{\lambda_0(\tau)}{N}
\end{equation}
Similarity we get for $[r_{av}^2]_{\tau,t_W}$ and $[r_M\cdot r_{av}]_{\tau,t_W}$
\begin{equation} \mylabel{a2}
[r_{av}^2]_{\tau,t_W}=\frac{1}{N}\sum_{\mu =0}^{N-1}a_\mu^2(\tau)\lambda_\mu(\tau)
\end{equation}
with
\begin{equation} \mylabel{a3}
a_\mu(\tau)=\frac{1}{\sqrt{N}}\sum_ie_i^\mu(\tau)
\end{equation}
$a_0$ corresponds to the mean value $\bar{\beta}$ of $\beta_i$ over $i$.
\begin{equation} \mylabel{a4}
[r_M\cdot r_{av}]_{\tau,t_W}=\frac{\lambda_0(\tau)}{N}\cdot \bar{\beta}
\end{equation}
Inserting equations \myref{a1}, \myref{a2} and \myref{a4} into $\Delta^2$ from
\myref{24} we get
\begin{equation} \mylabel{a5}
\Delta^2=(1-\bar{\beta})^2 + \sum_{\mu >0}\frac{\lambda_\mu(\tau)}
          {\lambda_0(\tau)}a_\mu^2(\tau)
\end{equation}
Since $e_i^\mu$ and $e_i^0$ are orthogonal for $\mu >0$ we can write
$a_\mu$ as
\begin{equation} \mylabel{a6}
a_\mu(\tau)=\frac{1}{\sqrt{N}}\sum_i e_i^\mu(1-\sqrt{N}e_i^0)
\end{equation}
Applying the Schwartz inequality to \myref{a6} we get
\begin{equation} \mylabel{a7}
a_\mu^2(\tau)\le 2(1-\bar{\beta})
\end{equation}
Together with $\sum_{\mu =0}\lambda_\mu=trace(C)$ this leads to the 
inequality \myref{29c} for $\Delta^2$. Since the average $C_{av}$ is a
function $[r_{av}^2]_t$ insertion of $\Delta^2$ into equation \myref{a2}
leads to equation \myref{29e}.

\section{\label{app2}Perturbation Expansion}
The matrix $C$ in equation \myref{32} is a sum of two matrices. The first
$C^0_{ij}=\beta^0_i\beta^0_j\gamma_M^2$ has one large eigenvalue 
$E_0=\gamma_M^2N$ with an eigenvector $f^0_i=\beta^0_i/{\sqrt{N}}$ 
and $N-1$ degenerate zero eigenvalues with vectors $f^\mu_i$
with $\mu>0$.These must satisfy only the orthogonality relation
\begin{equation} \mylabel{a8a}
(f^0,f^\mu)=0
\end{equation}
with ($a,b)$ denoting the scalar product.
To obtain a complete basis we impose on $f^\mu$ in the $N-1$ dimensional subspace
 the following conditions with the second matrix $C^1_{ij}=\delta_{ij}\gamma_i^2$
\begin{equation} \mylabel{a8}
(f^\nu,C^1\; f^\mu)=0 \quad \mbox{for} \; \mu\ne \nu \mbox{ and}\; \mu , \nu\ne\; 0
\end{equation}
We apply standard second order Rayleigh Schr\"odinger perturbation theory \footnote{See any textbook on quantum mechanics, e.g., \cite{quan}.}
with $C^1$ as perturbation. For general matrices $C^0$ and $C^1$ using
only the spectrum of $C^0$ and condition \myref{a8a} we get up to order 
$1/E_0^2$ for the leading eigenvalue
\begin{equation} \mylabel{a9}
\lambda_0=E_0+(f^0,C^1f^0)+\frac{1}{E_0}[(f^0,(C^1)^2f^0)-(f^0,C^1f^0)^2]
\end{equation}
and its eigenvector
\begin{equation} \mylabel{a10}
e^0_i=f^0_i+\frac{1}{E_0}[(C^1f^0)_i-(f^0,C^1f^0)f^0_i]
\end{equation}
The other eigenvalues require the in general complicated solution of
equation \myref{32} for $f^\mu_j$. They are given by
\begin{equation} \mylabel{a11}
\lambda_\nu=(f^\nu,C^1\;f^\nu)-\frac{1}{E_0}(f^0,C^1f^\nu)^2
\end{equation}
Inserting the specific form of $C^1$ we obtain for $\lambda_0$ and $\beta_i$ with
\begin{equation} \mylabel{a13}
(f^0,C^1f^0)=\frac{1}{N}\sum_i(\beta^0_i)^2\gamma_i^2=\langle \gamma^2\rangle_\beta
\end{equation} 
\begin{equation} \mylabel{a13a}
(f^0,(C^1)^2f^0)=\frac{1}{N}\sum_i(\beta^0_i)^4\gamma_i^2=\langle \gamma^4\rangle_\beta
\end{equation}
\begin{equation} \mylabel{a13b}
\lambda_0=\gamma_M^2N+\langle \gamma^2 \rangle_\beta +\frac{1}{\gamma_M^2N}
          \left[\langle \gamma^4 \rangle_\beta -(\langle \gamma^2 \rangle_\beta)^2 \right]
\end{equation}
\begin{equation} \mylabel{a14}
\beta_i=\beta^0_i \left(1+\frac{1}{\gamma_M^2N} \left(\gamma^2_i-\langle \gamma^2 \rangle \right) \right)
\end{equation}
$\langle \;\rangle_\beta$ denotes the average over $i$ weighted with $(\beta^0_i)^2$. 
Since neglected terms
are of order $1/N^2$ these formulae describe $\lambda_0$ and $\beta_i$
fairly accurate already for moderate $N$. 
Due to the degeneracy the general formalism of \cite{marc} for the modification 
due to noise does not apply. Using the \cite{wish} formula we find the
relative error in $\lambda_0$ due to a finite observation window $T$ is of
order $1/\sqrt{T}$ instead of $\sqrt{N/T}$ expected from \cite{marc} .
The non-leading eigenvalue will be changed considerably if the spread of
$\gamma_i$ is small, see \cite{burd}.

\section{\label{randomize} Significance of the order parameter}

\begin{figure}[ht]
   \includegraphics[width= \textwidth, clip=true]{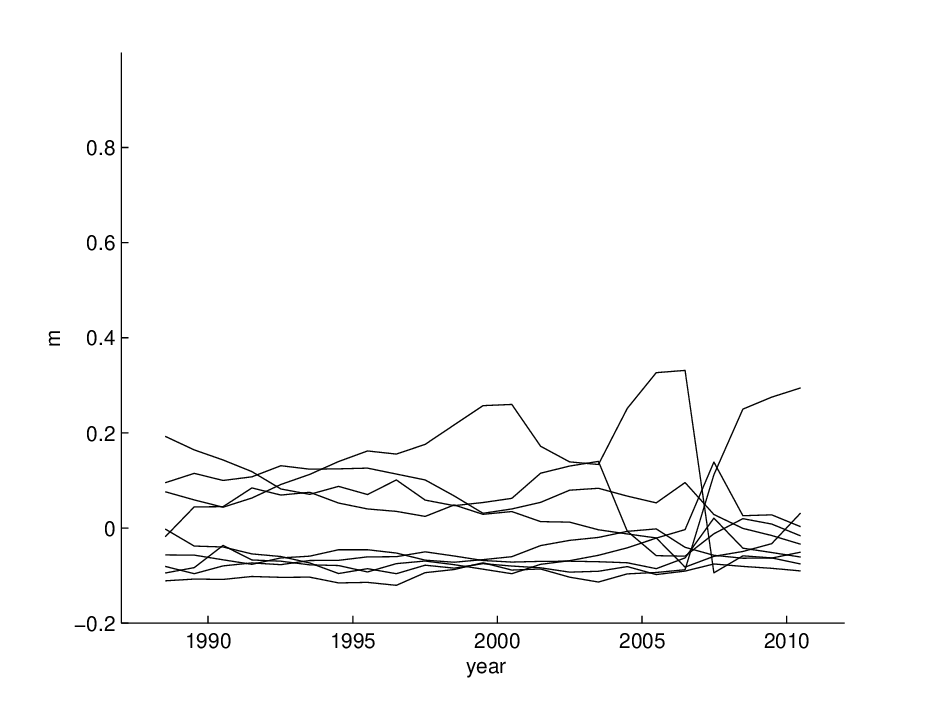}
   \caption{\label{fig:randomize} \em The risk parameter $m(\tau,s)$ with 
            the same $\beta_i$
            and $v_i$ as in eq. \myref{40a} but for shuffled sector affiliations $s$.}
\end{figure}

We analyzed the significance of the order parameter by assigning a random value for the sector to each stock. In figure \ref{fig:randomize} 
we show one representative result of such a random assignment. 
Compared with figure \ref{fig5}, the
large values of $m$ disappear. Before and after 2006 $m$ behaves similar as in a 
disordered phase. The structure observed near 2006 cannot be removed by re-shuffling,
because a substantial part of the stocks change their values of $\beta_i v_i$. If there 
would be only one stock in each IT or financial sector responsible for the transition 
seen in figure \ref{fig5}, one would get the same behavior. The fluctuation size of
$m$ in the order of $\pm 0.15$ indicates the error on the estimation of $m$. This
shows that the observed values for the IT and financial sector are far out
of the range of random fluctuations.

\end{appendix}

\end{document}